\documentclass[twocolumn,showpacs,superscriptaddress,nofootinbib]{revtex4-1}
\usepackage[T1]{fontenc} 
\usepackage{amsmath}
\usepackage{amssymb}
\usepackage{amsfonts}
\usepackage{graphicx}
\usepackage{enumitem}
\usepackage{bm}
\usepackage{rotating}
\usepackage{hyperref}
\usepackage{xcolor}
\usepackage{array}
\usepackage{lmodern}
\usepackage[normalem]{ulem}
\usepackage{braket}
\usepackage{tensor}
\usepackage{mathrsfs}
\usepackage{float}

\newcommand{\be}{\begin{equation}}
\newcommand{\ee}{\end{equation}}
\newcommand{\ba}{\begin{eqnarray}}
\newcommand{\ea}{\end{eqnarray}}
\newcommand{\beq}{\begin{equation}}
\newcommand{\eeq}{\end{equation}}
\newcommand{\beqa}{\begin{eqnarray}}
\newcommand{\eeqa}{\end{eqnarray}}
\newcommand{\nn}{\nonumber}
\newcommand{\sx}{\mathsf{x}}

\usepackage[normalem]{ulem}

\begin{document}

\title{Quantum Distinction of Inertial Frames: Local vs. Global}

\author{Wan Cong}
\email{wcong@uwaterloo.ca}
\address{Department of Physics and Astronomy, University of Waterloo, Waterloo,
Ontario, Canada, N2L 3G1}
\address{Perimeter Institute, 31 Caroline St., Waterloo, Ontario, N2L 2Y5, Canada}

\author{Ji\v r\'i Bi\v c\'ak}
\email{bicak@mbox.troja.mff.cuni.cz }
\address{Institute of Theoretical Physics, Faculty of Mathematics and Physics, V Hole\v sovi\v ck\'ach 2, 180 00 Prague 8, Czech Republic}

\author{David Kubiz\v n\'ak}
\email{dkubiznak@perimeterinstitute.ca}
\address{Perimeter Institute, 31 Caroline St., Waterloo, Ontario, N2L 2Y5, Canada}
\address{Department of Physics and Astronomy, University of Waterloo, Waterloo,
Ontario, Canada, N2L 3G1}

\author{Robert B. Mann}
\email{rbmann@uwaterloo.ca}
\address{Department of Physics and Astronomy, University of Waterloo, Waterloo,
Ontario, Canada, N2L 3G1}
\address{Perimeter Institute, 31 Caroline St., Waterloo, Ontario, N2L 2Y5, Canada}

\begin{abstract}
We study the response function of Unruh-deWitt detectors placed in a flat spacetime inside a thin matter shell.
We show that the response function distinguishes between the local and global (Minkowski) inertial frames and picks up the presence of the shell even when the detector is switched on for a finite time interval within which a light signal cannot travel to the shell and back as required by a classical measurement. We also analyze how the response of the detector depends on its location within the shell.
\end{abstract}

\date{March 11, 2020}

\maketitle

\section{Introduction}

Discerning the structure of spacetime is a task that is classically performed with clocks and rulers.  One locally sets up (in principle, if not in practice) a grid of rulers and an array of synchronized clocks, and then performs measurements on a variety of test bodies to determine their behaviour and from this infer the metric and curvature of spacetime in one's vicinity.

This procedure has limitations, one of the most notable being the distinction of local from global flatness.  An observer located inside a uniform static spherical shell does not experience any gravitational field.  All local measurements will indicate that test bodies move on straight-line geodesics.   However, this is the same result that would be obtained if the shell were absent, i.e., in a globally flat spacetime.  The only way this observer could classically detect the presence of the shell would be by sending out a probe and wait a sufficiently long time for the probe to hit the shell and send a signal back.  The minimal time required for the observer at the center of the shell to detect its presence would be the light-crossing time of the shell.

Recently it has been suggested that the situation is markedly different if one exploits quantum effects \cite{Ng:2016hzn}.  An observer making use of  an Unruh-deWitt (UdW) detector \cite{DeWitt:1980hx} (a 2-level qubit
that can be excited by a scalar field) can distinguish between the locally flat spacetime within the shell and globally flat Minkowski spacetime, even when it is effectively switched on (a Gaussian switching was used) for a time shorter  than the light-crossing time.  Even if the shell is  transparent
(i.e. non-interacting with the scalar field), having no net effect on the local gravitational field around the UdW detector, the non-local effects of gravity on the field vacuum yield a response in the detector that exhibits small but measureable differences from globally flat spacetime.

The reason for this is that the detector effectively exploits the fact that  the vacuum state of a quantum field contains information about global features of spacetime. This phenomenon has been seen in other contexts,
including vacuum entanglement harvesting
\cite{Valentini1991,Rez03,RRS05,Salton:2014jaa,Pozas2015,Pozas2016,Henderson:2017yuv,Ng:2018drz,Henderson:2018lcy}, probing topological features of spacetime that can induce preferred directions \cite{Martin-Martinez:2015qwa}, or that are even hidden behind event horizons \cite{Smith:2013zqa}.  Not only does the vacuum state of a quantum field carry non-local information about the \textit{gravitational} field -- a detector can read out such information locally.  Objects
in the vacuum likewise modify the mode structure of the vacuum in their vicinity, a feature that was recently exploited to demonstrate that a UdW detector can ``see in absolute darkness'', i.e., without exchanging any real quanta \cite{Ahmadzadegan:2019woa}.

Here we explore this phenomenon further,  demonstrating that a UdW detector that is causally disconnected from the external environment of the shell can still detect its presence.   We close a loophole present in previous work \cite{Ng:2016hzn,Ahmadzadegan:2019woa}, in which the interaction between the UdW detector and the quantum field had a Gaussian profile $\chi_G(\tau) = e^{-\tau^2/(2\sigma^2)}$ where $\tau$ is the proper time of the detector. This profile ensures that the UdW/field interaction never really drops to zero outside some finite time interval but instead persists for an infinitely long time, albeit being suppressed at large $\tau$.  We consider instead  a smooth, compact interaction profile for the detector, and place the detector at different radial positions within the shell.
 Our results not only confirm  previous studies \cite{Ng:2016hzn}, they strengthen them by causally isolating the detector from the shell during the time it is `switched on', which is for  a time interval that is shorter than the light travel time across the shell.

\section{Set-up}

In this section, we briefly review the set-up of the problem as presented in \cite{Ng:2016hzn}.
\subsection{Scalar Field Solution}
The shell spacetime is obtained by ``gluing'' together the Schwarzchild spacetime outside the shell with flat spacetime inside the shell, with the metric being at $r<R$
\begin{equation}
    ds^2=
      -f(R)dt^2+dr^2+r^2(d\theta^2+\sin^2\theta d\phi^2),
 \end{equation}
and
\begin{equation}
    ds^2=
      -f(r)dt^2+\frac{1}{f(r)}dr^2+r^2(d\theta^2+\sin^2\theta d\phi^2),
 \end{equation}
at $r>R$, where $f(r) = 1-2M/r$, with $M$ being the mass and $R$ the radius of the shell. As shown in \cite{Ng:2016hzn}, this metric satisfies the two junction conditions (see, for example, Section 3.7 of \cite{poisson_2004}) needed for the spacetime to be a well-defined solution to the Einstein field equations.  The shell is massive, exhibiting a spherically symmetric gravitational field outside, but does not interact with the scalar field -- it
is transparent to scalar matter.

Using this metric the massless scalar field equation,
\begin{equation}
    \partial_{\mu}(\sqrt{-g}g^{\mu\nu}\partial_{\nu}\Psi) = 0\,,
\end{equation}
admits the usual separable solutions of the form
\begin{equation}
\label{sep}
    \Psi_{\omega\ell m}(t,r,\theta,\phi) = \frac{1}{\sqrt{4 \pi \omega}}e^{-i\omega t}Y_{m\ell}(\theta,\phi)\psi_{\omega\ell}(r)\,.
\end{equation}
In the above, the mode indices $\omega\in (0,\infty)$, $\ell\in \mathcal{Z}$, $m = -\ell, -\ell+1,...,\ell-1,\ell$ and $Y_{m\ell}$ are the spherical harmonics normalised as
$$
\int_{S^2} Y^*_{m_1\ell_1}Y_{m_2\ell_2} dA= \delta_{m_1,m_2}\delta_{\ell_1,\ell_2} \;  .
$$

 The resulting radial equation for $\psi_{\omega\ell}(r)$ is
\begin{align}
    \label{eq: radialKG}
&\omega^2 \psi_{\omega\ell}(r)+\frac{\alpha}{\beta \,r^2}\frac{d}{dr}\Big(\frac{\alpha r^2}{\beta} \frac{d}{dr}\psi_{\omega\ell}(r)\Big) \nonumber\\
&\qquad \qquad\qquad -\bigg(\frac{\alpha^2\ell(\ell+1)}{r^2}\bigg)\psi_{\omega\ell}(r) =0,
\end{align}
where the functions $\alpha$ and $\beta$ are
\begin{align}
\label{eq: ab}
    \alpha(r) &= \begin{cases}
      \sqrt{f(R)}, & r\leq R \\
      \sqrt{f(r)}, & r> R \\
   \end{cases}\,, \nonumber \\  \beta(r) &= \begin{cases}
      1, & r\leq R \\
      1/\sqrt{f(r)}, & r> R\,. \\
   \end{cases}
\end{align}
Inside the shell the radial equation explicitly reads
\begin{equation}
   \frac{\omega^2}{f(R)}r^2\psi_{\omega\ell} + 2r \frac{d}{d r}\psi_{\omega\ell} +r^2\frac{d^2}{dr^2}\psi_{\omega\ell} -\ell(\ell+1)\psi=0,
\end{equation}
whose solutions are spherical Bessel functions of the first kind, $j_{\ell}(\tilde\omega r)$, where $\tilde\omega=\frac{\omega}{\sqrt{f(R)}}$. The solution outside the shell has to be determined numerically and matched to the solution on the shell.

Continuity of the radial solution is imposed at the boundary of the shell by setting $\psi_{\omega\ell}(R) = j_{\ell}(\tilde\omega R)$. This in turn fixes the jump in the derivatives of $\psi_{\omega\ell}$ across the shell. To find the value of $d\psi_{\omega\ell}/dr|_{R^+}$, we integrate Eq.~\eqref{eq: radialKG} across the shell, obtaining the condition
\begin{equation}
    \bigg[\frac{\alpha(r)}{\beta(r)}\frac{d}{dr}\psi_{\omega\ell}\bigg]=0\,,
\end{equation}
where the square brackets represent the difference in the value of the term across the shell. Noting the discontinuity in the coefficient $\beta(r)$ across the shell, \eqref{eq: ab}, this yields the desired initial conditions $\psi_{\omega\ell}(R^+)$ and $\psi_{\omega\ell}'(R^+)$ for  numerically solving the radial equation outside the shell.

Finally, to normalize the solution, we will follow the scheme presented in \cite{Ng:2016hzn}. First, the radial equation~\eqref{eq: radialKG} for $r>R$ can be rewritten in terms of a new coordinate $r^{\star}$ such that $d/dr^{\star} = \frac{\alpha}{\beta}d/dr$. Further, defining $\rho_{\omega\ell} = r \psi_{\omega\ell}$, the radial equation reads
\begin{equation}
   \frac{d^2}{dr^{\star^2}}\rho_{\omega\ell}+(\omega^2-V(r))\rho_{\omega\ell}=0\,,
\end{equation}
where
\begin{equation}\label{Veff}
    V(r) = \frac{\alpha^2\ell(\ell+1)}{r^2}+\frac{1}{r}\frac{\alpha}{\beta}\frac{d}{dr}\bigg(\frac{\alpha}{\beta}\bigg).
\end{equation}
Asymptotically, $V(r)\rightarrow 0$ as $r\rightarrow \infty$ and hence $\psi_{\omega\ell} \sim \sin(\omega r^{\star})/r^{\star}$. Let the normalized radial solution be written as $\psi_{\omega\ell}(r^{\star})=A_{\omega\ell}\tilde\psi_{\omega\ell}(r^{\star})$. Given any two wavefunctions $\Psi_1,\,\Psi_2$, the Klein-Gordon inner product between them is
\begin{equation}
    (\Psi_1,\Psi_2) = i\int_{\Sigma} d\sigma n^{\mu}(\Psi_1^{*}\nabla_{\mu}\Psi_2-\Psi_2\nabla_{\mu}\Psi_1^{*})\,,
\end{equation}
where $\Sigma$ is a Cauchy surface with normal $n^{\mu}$. The solution~\eqref{sep} will be normalized with respect to the Klein-Gordon inner product if we choose the normalisation constant $A_{\omega\ell}$ such that $A_{\omega\ell}\tilde\psi_{\omega\ell}(r^{\star})\rightarrow 2\sin(\omega r^{\star})/r^{\star}$ as $r^{\star}\rightarrow \infty$ \cite{Ng:2016hzn}. This fixes the boundary condition at infinity for the determination of $A_{\omega\ell}$. Meanwhile, the normalisation constant in the Minkowski case is $2\omega$, giving the full normalised solution, $\Psi^M_{\omega\ell m}(t,r,\theta,\phi) = \sqrt{\frac{\omega}{\pi}}e^{-i\omega t}Y_{m\ell}(\theta,\phi)j_{\ell}(\omega r)$ \cite{Ng:2016hzn}.

\subsection{UdW response}
\label{sec: 2b}
\begin{figure}[t]
    \centering
    \includegraphics[scale=0.4]{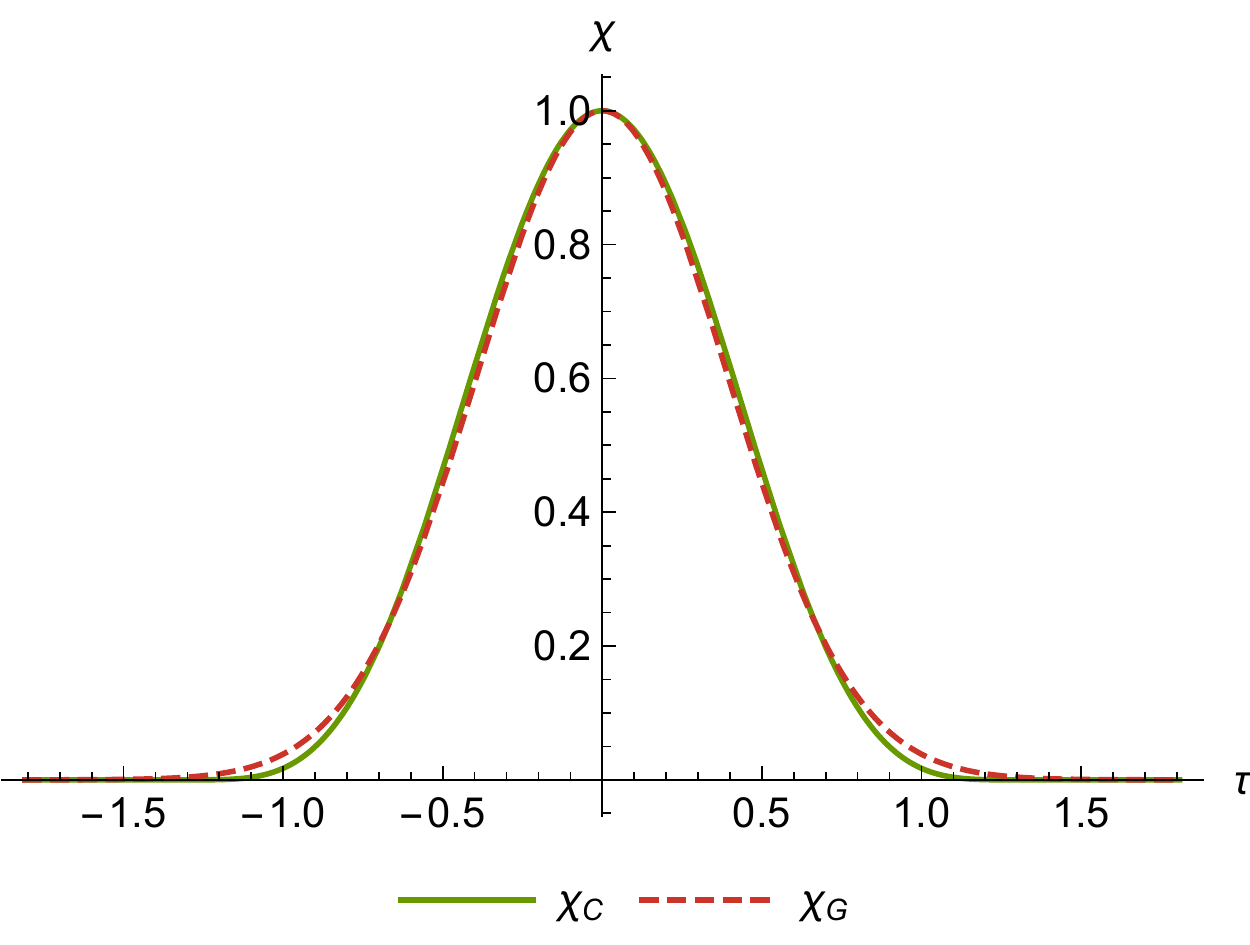}
    \caption{{\bf Plot of Gaussian $\chi_G$ and $\chi_c$ switching.} Here, the parameters used are $\eta=1.2,\,\sigma=\frac{3}{8\eta}\sqrt{\frac{\pi}{2}}$.  These parameters give the same area under the graph for the two switching profiles. Although both look similar, $\chi_c$ is compactly supported. The interaction duration between the detector and the quantum field can thus be made truly shorter than the light-crossing time of the shell.}
    \label{fig:svf}
\end{figure}

A UdW detector is a 2-level quantum mechanical system that interacts locally with a scalar quantum field $\hat{\phi}$ as it moves along some trajectory $\sx(\tau)$ in spacetime. Let $\Omega$ denote the energy gap of the detector and
$$
\hat{\mu}(\tau)=e^{-i\Omega\tau}\hat\sigma^++e^{i\Omega\tau}\hat\sigma^-
$$
its monopole moment  (in the interaction picture), where $\hat{\sigma}^{\pm}$ are the ladder operators. The Hamiltonian governing the detector/field  interaction reads
\begin{equation}
     \hat H(\tau) = \lambda\chi(\tau)\hat\mu(\tau)\otimes\hat\phi(\sx(\tau))\,,
\end{equation}
where $\tau$ is the proper time of the detector and $\lambda$ is the dimensionless coupling constant.
We choose the switching function $\chi_c(\tau)$ to be
\begin{align}
    \chi_c(\tau) &= \begin{cases}
      \cos^4(\eta \tau), &-\frac{\pi}{2 \eta}\leq \tau\leq\frac{\pi}{2 \eta}\\
       0, &\text{otherwise}
   \end{cases}\label{eq:compact}
\end{align}
so as to ensure a finite duration of interaction.
Thus the interaction switches on and off smoothly and takes place between the finite time interval $\tau\in(-\frac{\pi}{2 \eta},\frac{\pi}{2 \eta})$ for some $\eta>0$. We have chosen this particular form of the switching because it has a shape similar to the Gaussian switching function $\chi_G$ used in \cite{Ng:2016hzn} (see Fig. \ref{fig:svf}).

If the detector starts off in the ground state and interacts with the quantum vacuum via the above Hamiltonian, there may be a non-zero probability of finding the detector in its excited state after the interaction. The probability of excitation of the detector can be calculated using perturbation theory and is well-known in the literature. It is given by \cite{birrell_davies_1982,Pozas2015}
\begin{equation}
\begin{split}
    P = \lambda^2&\int_{-\infty}^{\infty}  d \tau_1\,\int_{-\infty}^{\infty}d \tau_1 \chi_c(\tau_1)\chi_c(\tau_2) e^{-i\Omega(\tau_2-\tau_1)}\\&\times W(\sx(\tau_1),\sx(\tau_2))
    \end{split}
\end{equation}
to second order in $\lambda$, where $W(\sx(\tau_1),\sx(\tau_2))$ is the Wightman function of the field evaluated along the detector trajectory.

The field operator can be expanded in terms of the normalized field modes $\Psi_{\omega\ell m}$ as
\begin{multline}
    \hat\phi(\sx(\tau)) =\sum_{\ell,m}\int_0^{\infty}d\omega\, \hat{a}_{\omega\ell m}\Psi_{\omega\ell m}(\sx(\tau))\\+\hat{a}^{\dagger}_{\omega\ell m}\Psi^{\dagger}_{\omega\ell m}(\sx(\tau))\,,
\end{multline}
with $\hat{a}_{\omega\ell m}$ denoting the mode annihilation operators. Let $\ket{0}$ denote the field vacuum such that $\hat{a}_{\omega\ell m}\ket{0}=0$. This corresponds to the vacuum with respect to an observer located at infinity.  We note that this vacuum also corresponds to that of an inertial observer inside the shell, since the mode solutions in Eq.~\eqref{sep} are positive frequency with respect to the proper times of both these observers -- the Bogoliubov transformation between the inside and outside modes does not mix creation and annihilation operators.

The Wightman function with respect to this vacuum $W(\sx(\tau_1),\sx(\tau_2)) := \bra{0}\hat\phi(\sx(\tau_2))\hat\phi(\sx(\tau_1))\ket{0}$ is given by
\begin{equation}
     W(\sx(\tau_1),\sx(\tau_2))=\sum_{\ell,m}\int_0^{\infty}d\omega \Psi^{\dagger}_{\omega\ell m}(\sx(\tau_1))\Psi_{\omega\ell m}(\sx(\tau_2))\,.
\end{equation}

From the previous section, we have seen that the normalized mode solutions are given by $\Psi_{\omega\ell m} = \frac{1}{\sqrt{4\pi\omega}}e^{-i \omega t}Y_{\ell m}(\theta,\phi)A_{\omega\ell}j_{\ell}(\tilde\omega r)$ inside the shell. We are interested in studying how the response of the detector differs when placed respectively in a spherical shell and globally flat Minkowski spacetime. A simple choice for the trajectory $\sx(\tau)$ of the detector is $r=r_d$, $\theta=\pi/2$, $\phi = 0$. In this case, noting that $t=\tau/\sqrt{f(R)}$, we have
\begin{widetext}
\begin{align}\label{eq: responsei}
   \nn \mathcal{F}&= \int_{-\infty}^{\infty}  d \tau_1\,\int_{-\infty}^{\infty}d \tau_2 \chi_c  (\tau_1)\chi_c(\tau_2) e^{-i\Omega(\tau_2-\tau_1)}\sum_{\ell m}\int_0^{\infty}d\omega\Psi^{\dagger}_{\omega\ell m}(\sx(\tau_1))\Psi_{\omega\ell m}(\sx(\tau_2))\\ \nn
   &=\int_{-\infty}^{\infty}  d \tau_1\,\int_{-\infty}^{\infty}d \tau_2 \chi_c(\tau_1)\chi_c(\tau_2) e^{-i\Omega(\tau_2-\tau_1)}\sum_{\ell m}\int_0^{\infty}\frac{d\omega}{4\pi\omega}e^{-i\tilde\omega(\tau_2-\tau_1)}|Y_{\ell m}(\frac{\pi}{2},0)|^2|A_{\omega\ell}|^2 |j_{\ell}(\tilde\omega\, r_d)|^2\\
   &=\sum_{\ell m}\int_0^{\infty}
\frac{d\omega}{4\pi\omega}\int_{-\infty}^{\infty}  d \tau_1\,\int_{-\infty}^{\infty}d \tau_2 \chi_c(\tau_1)\chi_c(\tau_2) e^{-i(\Omega+\tilde\omega)(\tau_2-\tau_1)}|Y_{\ell m}(\frac{\pi}{2},0)|^2|A_{\omega\ell}|^2 |j_{\ell}(\tilde\omega\,  r_d)|^2\,,
\end{align}
\end{widetext}
for  the \textit{response function}  $\mathcal{F}=P/\lambda^2$ of the field,
where in the last step we switched the order of integration since the integrand is smooth. This expression can be further simplified by integrating over the $\tau_1$ and $\tau_2$ variables, which amounts to performing Fourier transforms on the switching functions. Denoting the Fourier transform of the switching function as
\begin{equation}
    \hat\chi_c(k)=\frac{1}{\sqrt{2\pi}}\int_{-\infty}^{\infty}d\tau\chi_c(\tau)e^{-ik\tau}\,,
\end{equation}
the response function~\eqref{eq: responsei} simplifies to

\begin{widetext}
\begin{align}\label{eq: response}
\mathcal{F}&=\sum_{\ell m}\int_0^{\infty}
\frac{d\omega}{2\omega} \left|\hat{\chi}_c  \left(\Omega+\tilde\omega\right)\right|^2|A_{\omega\ell}|^2|Y_{\ell m}(\frac{\pi}{2},0)|^2|j_{\ell}(\tilde\omega r_d)|^2\,,
\end{align}
\end{widetext}
upon using the fact that $\hat{\chi}_c(-k)=\hat{\chi}_c(k)$ for a real switching function. Explicitly, we have
\begin{equation}
    \label{eq: FT}
    \hat\chi_c(k) =\sqrt{\frac{2}{\pi}} \frac{24 \eta^4\sin\frac{\pi k}{2\eta}}{64\eta^4k-20\eta^2 k^3+k^5}\,.
\end{equation}

\section{Results}

We are now ready to study how the presence of the shell affects the response of a UdW detector inside a shell compared to its response in globally flat Minkowski space.
\begin{figure}[t]
    \centering
    \includegraphics[scale=0.5]{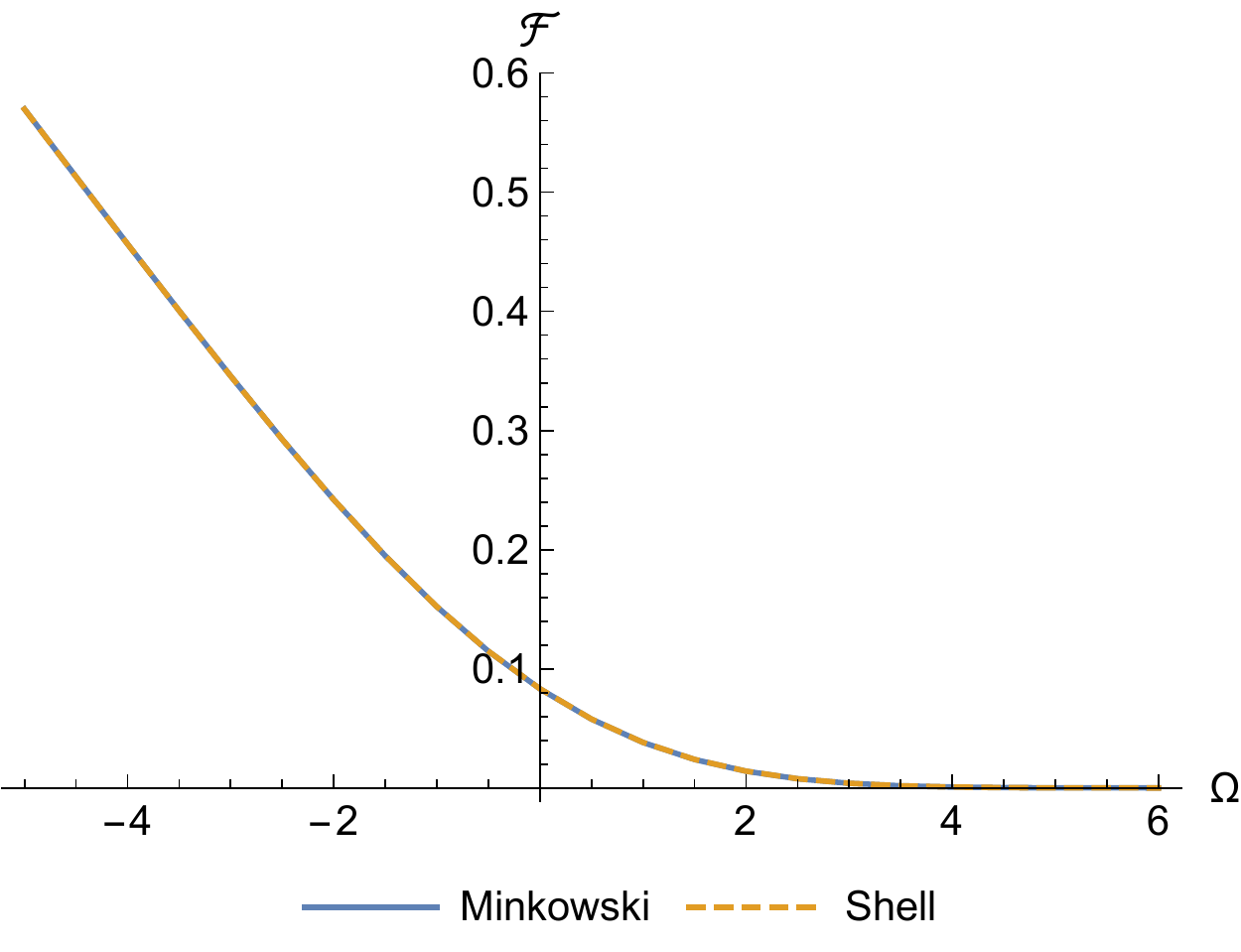}
    \includegraphics[scale=0.5]{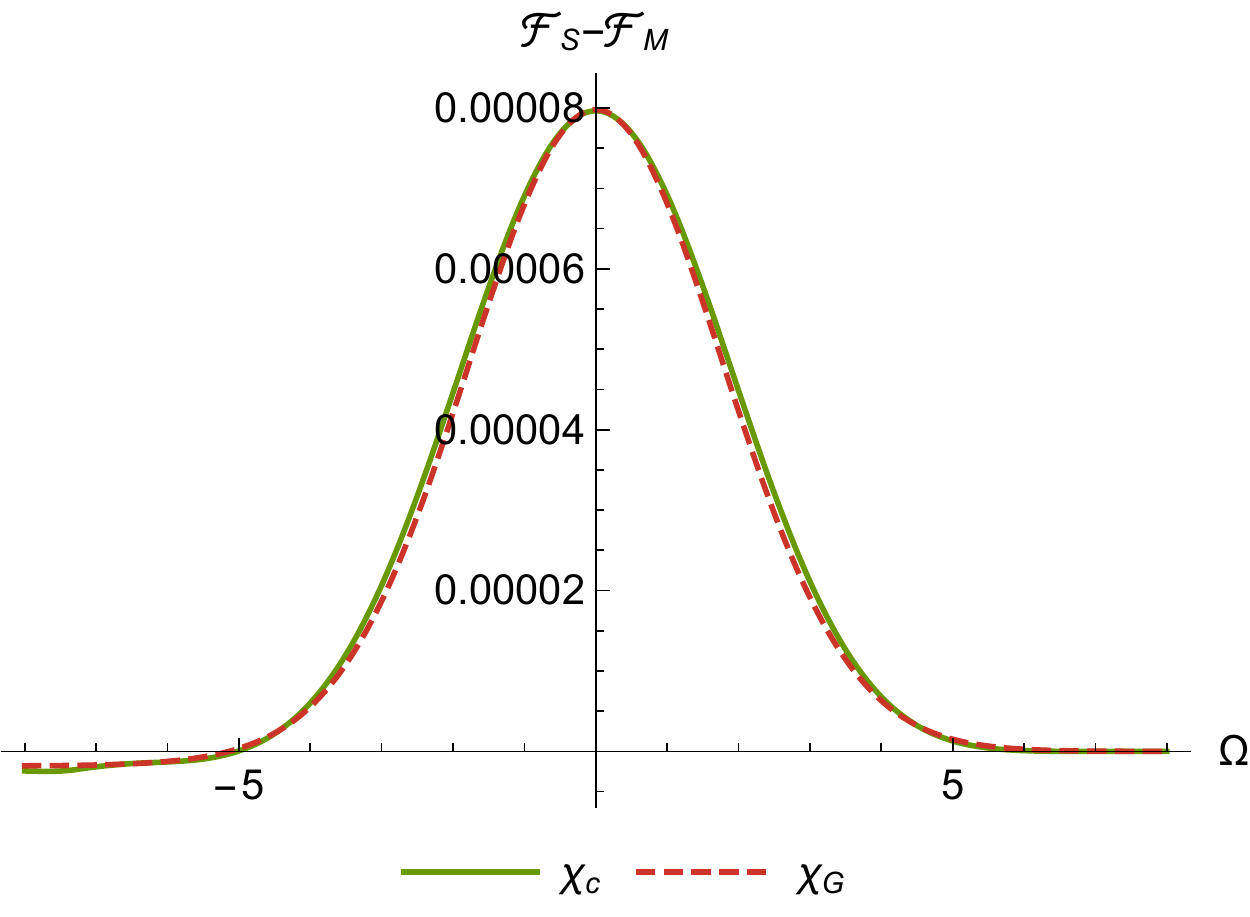}
    \caption{{\bf Detector response against $\Omega$.} {\em Top:} Plot of $\mathcal{F}$ against $\Omega$ for both the shell (yellow) and globally flat Minkowksi spacetime (blue) for $M=0.5,\, R=3\,, \eta=1.2,\,r_d=0$. The two cases are indistinguishable on the scale of this figure, but the differences can be studied by looking at the bottom figure. {\em Bottom:} Plot of the difference $\mathcal{F}_S-\mathcal{F}_M$ against $\Omega$. The results obtained using $\chi_c$ and $\chi_G$ ($\sigma=\frac{3}{8\eta}\sqrt{\frac{\pi}{2}}$) are qualitatively similar.}
    \label{fig: Fvgap}
\end{figure}
We find small but significant differences between the two cases, using $\mathcal{F}_S$ to denote
the response of a detector placed in a spherical shell, and $\mathcal{F}_M$ to denote its response in global Minkowski space.

Plotting $\mathcal{F}$ against $\Omega$ in Fig. \ref{fig: Fvgap}, we see that  the detector is indeed sensitive to the presence of the shell.  This is most apparent when we plot the difference $\mathcal{F}_S-\mathcal{F}_M$. Note that $\Omega < 0$ physically means that the detector starts off in its excited state. We have chosen the parameter $\eta$ such that interaction duration  $\pi/\eta\approx 2.6$ between the field and detector is less than $2R=6$, the time needed for a light signal to travel from the detector at the center to the shell and back. This is in contrast to the classical case, where the fastest way a detector inside the shell can detect its presence  is by sending and waiting for a light signal to come back from the shell. We thus strengthen the claim made in \cite{Ng:2016hzn}: a UdW interacting with the quantum vacuum can detect the shell faster than a classical detector even if its interaction time is causally disconnected from the shell.

\begin{figure}
    \centering
    \includegraphics[scale=0.55]{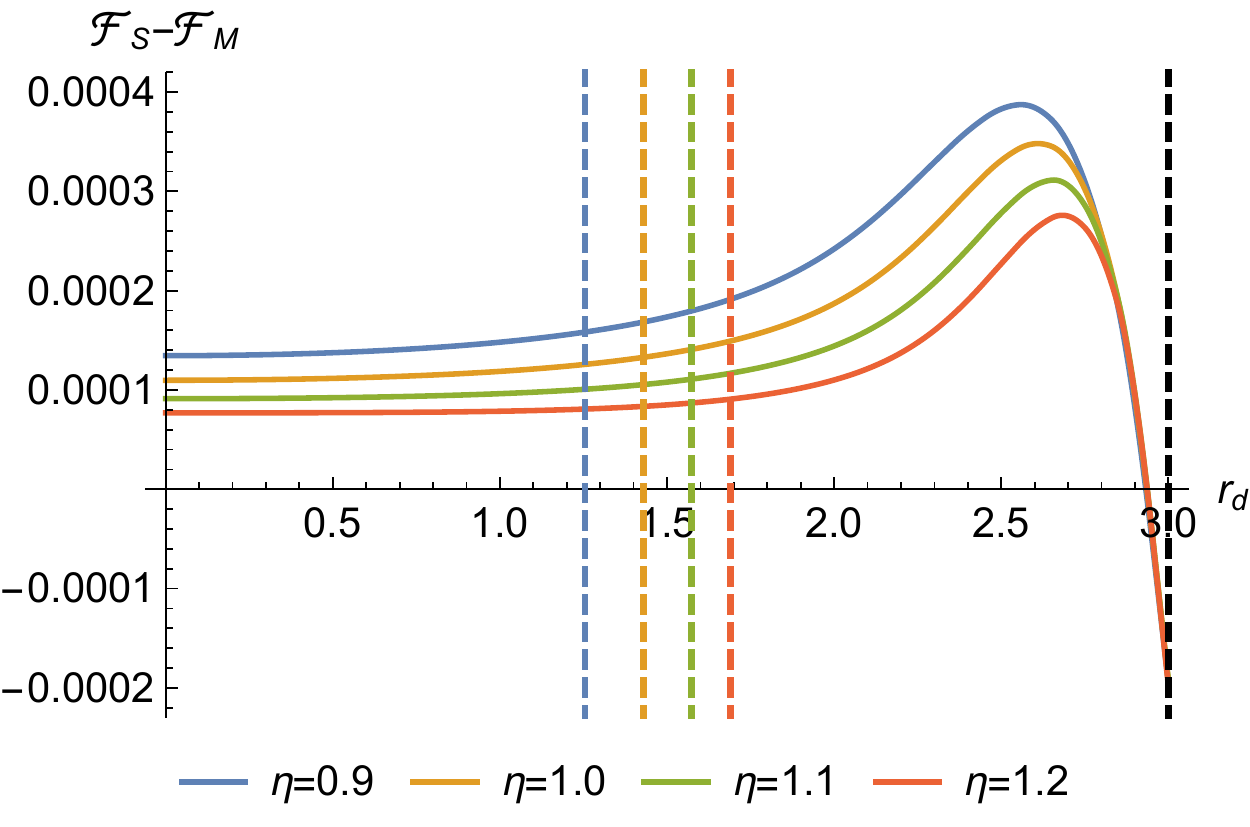}
    \caption{{\bf Plot of $\mathcal{F}_S-\mathcal{F}_M$ against $r_d$.} This plot is obtained by setting $\Omega=0.5\,,M=0.5,\, R=3\,\text{(black, dashed)}$ and $\eta=0.9,\,1,\,1.1,\,1.2$. The peaks indicate the optimal $r_d$ inside the shell at which the detector, for a given $\eta$, can best detect the presence of the shell.
The vertical dashed lines indicate, for a given $\eta$, the causal boundary of the interaction duration:  to the left of these lines this duration is less than the light-travel time across the shell.}
    \label{fig: radialdistance}
\end{figure}

We next consider the response of the detector as we vary its location $r_d$ within the shell.
Fig. \ref{fig: radialdistance} shows a plot of $\mathcal{F}_S-\mathcal{F}_M$ against $r_d$ with $\Omega=0.5$ and various choices of the interaction duration
$1/\eta$. For each $\eta$, as $r_d$ increases, the difference in response first decreases very slightly, before increasing to a peak lying between the left and right dashed lines in the figure. This can be interpreted as the existence of an optimal position at which the UdW detector can best detect the presence of the shell. However, at this position, the detector is switched on for a time longer than the light-crossing time. The largest $r_d$ beyond which this happens is indicated by the vertical dashed lines for each $\eta$.

\begin{figure}[t]
    \centering
    \includegraphics[scale=0.5]{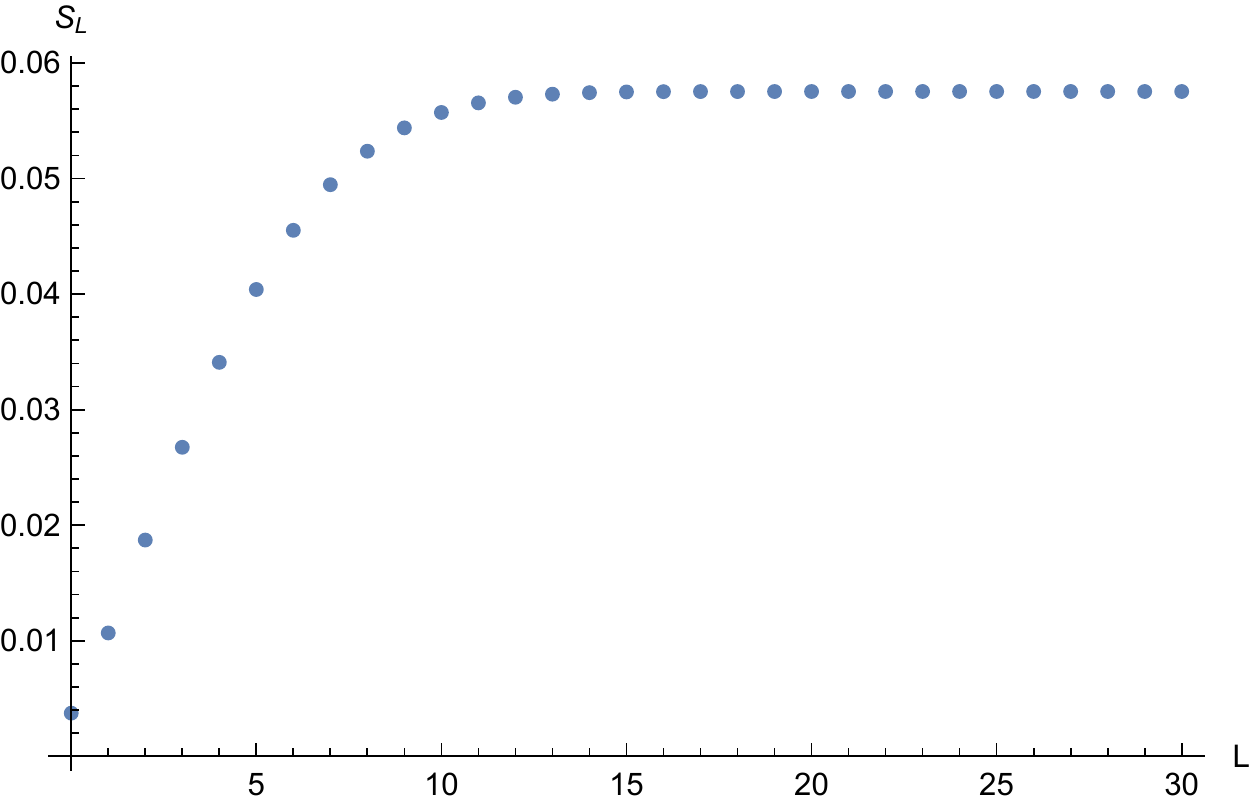}
    \vspace{0.1cm}
    \includegraphics[scale=0.5]{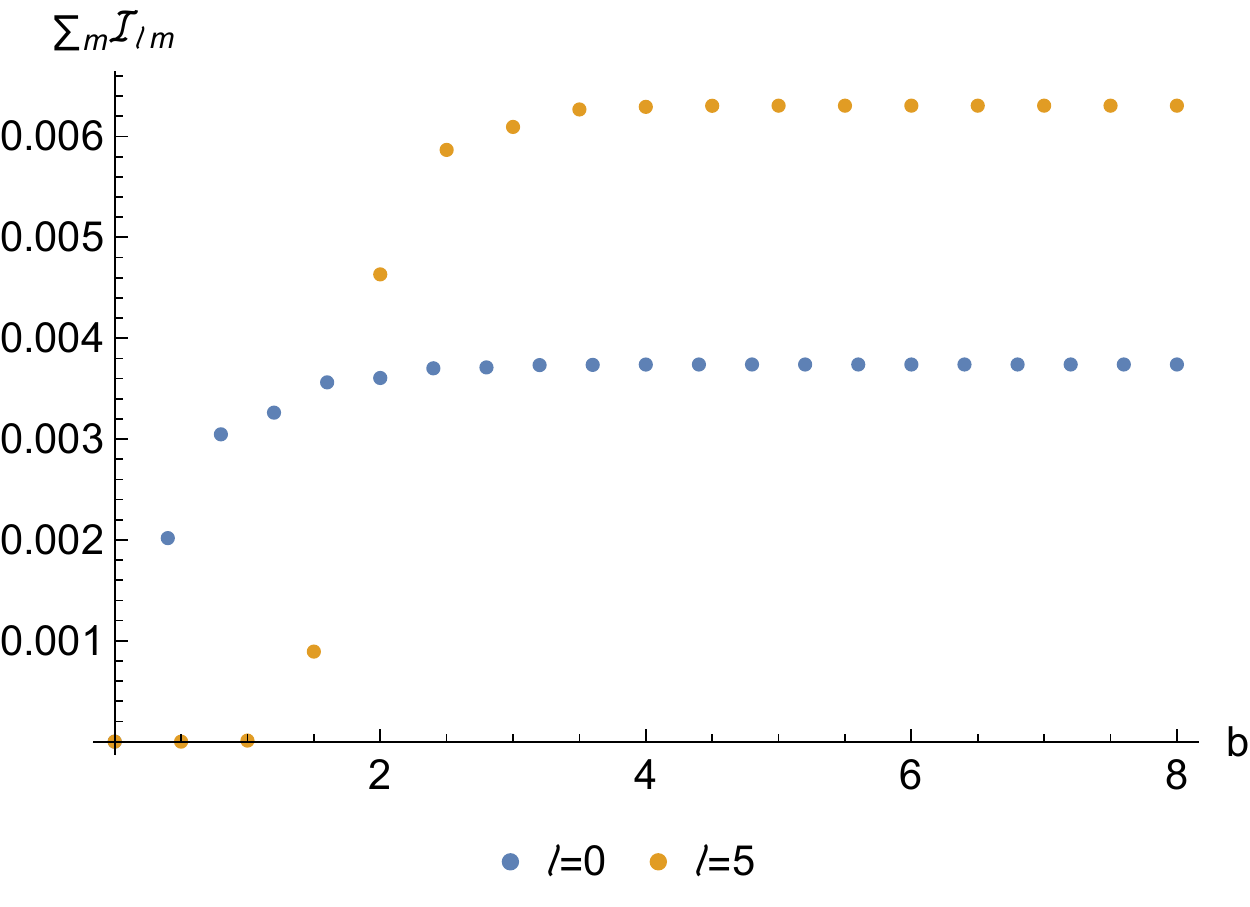}
    \caption{{\bf Numerical convergence.} In evaluating~\eqref{eq: response} numerically, an upper cut-off for the sum in $\ell$ has to be chosen. {\em Top:} Plot of the partial sum $\mathcal{S}_{L}$ against $L$. We see that the summation over $\ell$ is clearly convergent. The parameters used here are the same as those in Fig. \ref{fig: radialdistance}, with $r_d=3$. {\em Bottom:} Plot of $\sum_m \mathcal{I}_{\ell m}$ against the upper cut-off $b$. For each $\ell$ (two examples are shown here), the integral over $\omega$ is also clearly convergent. }
    \label{fig: convergence}
\end{figure}

We close this section by commenting on the  stability of our results, which  were computed by evaluating expression~\eqref{eq: response} numerically. In doing so, we have chosen upper cut-offs for the summation over $\ell$ and for the integral over $\omega$. Both the integral and summation exhibit clear numerical convergence, as shown in Fig. \ref{fig: convergence}, with
\begin{align}
\label{eq: partial sum}
S_L& =\sum_{\ell=0}^{L}\sum_{m=-\ell}^{\ell}\mathcal{I}_{\ell m}\,,\\
\mathcal{I}_{\ell m} &=
\int_0^{b}
\frac{d\omega }{2\omega} \left|\hat{\chi}_c  \left(\Omega+\tilde\omega\right)\right|^2|A_{\omega\ell}|^2|Y_{\ell m}(\frac{\pi}{2},0)|^2|j_{\ell}(\tilde\omega r_d)|^2\,.
\end{align}

For the results presented in this paper, we have chosen the cut-offs $L$ and $b$ such that the contribution of the next term in the summation or the next integral interval is less than $10^{-7}$.

\section{Conclusions}

The quantum vaccum affords much opportunity to explore the structure of spacetime in ways that are not possible classically. We have demonstrated that  a UdW detector that is causally disconnected from the external environment of the shell can still detect its presence relative to globally flat spacetime.  In so doing we have demonstrated a `quantum detection of local frame' phenomenon, in which non-local information about the global structure of spacetime contained in  the vacuum state of a quantum field can be read locally by a detector. 

A similar effect was also found in an idealized model in \cite{Lochan:2016cxt}, where it was shown that the in-vacuum in a $(1+1)$-dimensional Dilatonic black hole spacetime formed by a left-moving null wall has a thermal spectrum with respect to inertial observers located at the left side of the wall. Like the shell scenario considered in this paper, the spacetime around these observers is locally flat. A  UdW placed there will similarly record a difference in response depending on whether the wall is present or not. However the difference in response in this case is an Unruh-like effect, caused by non-trivial Bogoliubov transformation between the modes of the two vacua. In contrast, our UdW detector in the shell registers zero particle expectation number, as noted already in section \ref{sec: 2b}.

We have also shown that the detector can be  placed within the shell in different locations to optimally distinguish the local/global cases, but this optimal placement is not causally disconnected from the shell boundary.

We note that, although our work was carried out in the context of general relativity, its implications are considerably broader.  The $A_{\omega\ell}$ quantities depend on the form of the effective potential \eqref{Veff}, and thus upon the theory of gravity that governs the dynamics of spacetime.  In this sense a UdW detector is a non-local probe of the local dynamics of gravity outside of the shell.  A more complete study of this would be an interesting subject for future investigation.   

We can likewise ask if a  detector could be used to discern other effects, such as the dragging of inertial frames.  Work on this is in progress.
\bigskip

 \section*{Acknowledgments}
 \label{sc:acknowledgements}

 This work was supported in part by the Natural Sciences and Engineering Research Council of Canada and by the Perimeter Institute. J.B. thanks the kind hospitality of the Perimeter Institute, Waterloo, where this work started and the Albert Einstein Institute, Golm, where it continued. J.B. also acknowldeges the support from the Grant Agency of the Czech Republic, Grant No. GA\u CR 19-01850S.
 Research at Perimeter Institute is supported in part by the Government of Canada through the Department of Innovation, Science and Economic Development Canada and by the Province of Ontario through the Ministry of Colleges and Universities.


%

\end{document}